\documentclass[times, twoside]{zHenriquesLab-StyleBioRxiv-Preprint}
\usepackage{blindtext}
\usepackage{graphicx}
\usepackage{dblfloatfix} 
\usepackage{lineno}
\usepackage{comment}
\usepackage[normalem]{ulem}
\usepackage{color, colortbl}
\usepackage{xcolor}
\usepackage{MnSymbol}

\leadauthor{Willomitzer} 

\begin{document}

\pagecolor{white}

\title{Synthetic Wavelength Holography: \\ An Extension of Gabor's Holographic Principle to Imaging with Scattered Wavefronts}

\shorttitle{Synthetic Wavelength Holography}


\author[1,*]{Florian Willomitzer}
\author[2]{Prasanna V. Rangarajan}
\author[1]{Fengqiang Li} 
\author[2]{Muralidhar M. Balaji}
\author[2]{Marc P. Christensen}
\author[1]{Oliver~Cossairt}

\affil[1]{Department of Electrical and Computer Engineering, Northwestern University, Evanston, IL 60208}
\affil[2]{Department of Electrical and Computer Engineering, Southern Methodist University, Dallas, TX 75205}
\affil[ ]{}

\affil[*]{Correspondence: florian.willomitzer@northwestern.edu}

\maketitle

\begin{figure*}[t!]
\centering
\includegraphics[width=\linewidth]{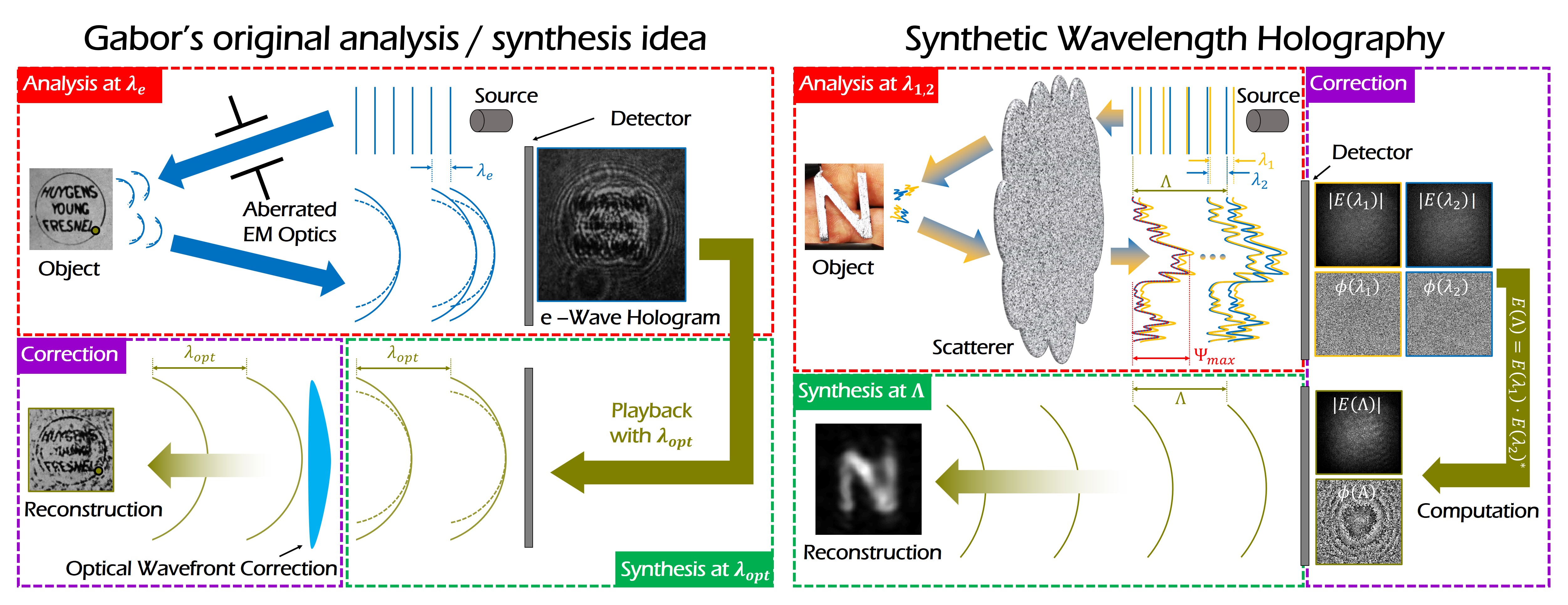}
\caption{Our method of 'Synthetic Wavelength Holography' is inspired by Gabor's idea of \textit{Analysis}, \textit{Synthesis}, and \textit{Correction} for improving the resolution of Electron Microscopes. Left: Gabor envisioned recording an electron wavefront with aberrated electron microscope optics (\textit{Analysis}, wavelength $\lambda_e$), then reconstructing this electron image by playing the hologram back with an optical wavefront (\textit{Synthesis}, wavelength $\lambda_{opt} >> \lambda_e$) while exploiting optical wavefront correction (\textit{Correction}, wavelength $\lambda_{opt}$) \cite{GaborNobel}. 
Right: In Synthetic Wavelength Holography, we adopt Gabor's initial idea to correct unknown wavefront aberrations $\Psi$ introduced when visible light is transported through scenes with strong scattering. We capture two holograms at two closely spaced wavelengths (\textit{Analysis}, wavelengths $\lambda_1$ and $\lambda_2$) each showing random aberrations. By computationally beating the two signals together, we produce a low frequency `Synthetic Wavelength Hologram' (\textit{Correction}). The Synthetic Wavelength Hologram is not subject to aberrations and contains information on the order of a `Synthetic Wavelength'  $\Lambda$, which is the beat wavelength of  $\lambda_1$ and $\lambda_2$. Similar to Gabor's idea, the object is reconstructed by playing back the computationally corrected hologram with the much larger Synthetic Wavelength $\Lambda$ (\textit{Synthesis}).}
\label{fig:Fig_AH}
\end{figure*}

\begin{abstract}

The presence of a scattering medium in the imaging path between an object and an observer is known to severely limit the visual acuity of the imaging system. We present an approach to circumvent the deleterious effects of scattering, by exploiting spectral correlations in scattered wavefronts. Our method draws inspiration from Gabor’s attempts to improve the resolving power of electron microscopes by recording aberrated wavefronts at electron wavelengths, followed by aberration correction and playback at optical wavelengths. We extend the notion to scattered wavefronts, by interpreting the scattering of light as a source of randomized aberration. We compensate for these aberrations by mixing speckle fields recorded at two closely spaced optical wavelengths $\lambda_1,\lambda_2$, and replaying the computationally assembled wavefront at a 'Synthetic Wavelength’ $\Lambda=\frac{\lambda_1\lambda_2}{|\lambda_1-\lambda_2|} ~>>\lambda_1,\lambda_2$. An attractive feature of our method is that it accommodates a wide variety of scattering mechanisms and operates at the physical limits of imaging in the presence of scatter. Moreover, our findings are applicable to other wave phenomena, opening up new avenues for imaging with scattered wavefronts.

\end {abstract}

\vspace{-3mm}

\section*{Introduction}

In his 1971 acceptance speech for the Nobel Prize in Physics, Denis Gabor spoke of the moment that led to his discovery of the holographic imaging principle:

\begin{quote}
\emph{    
"After pondering this problem for a long time, a solution suddenly dawned on me, one fine day at Easter 1947, ... Why not take a bad electron picture, but one which contains the whole information, and correct it by optical means? The electron microscope was to produce the ... interference pattern I called a `hologram', from the Greek word `holos' -the whole, because it contained the whole information. The hologram was then reconstructed with light, in an optical system which corrected the aberrations of the electron optics" \cite{GaborNobel}. }  
\end{quote}

Central to Gabor's award winning research were two innovative ideas. The first is the notion that an interferogram acquired at electron wavelengths provides a complete ('whole' or 3D) representation of atomic structure, warranting the designation of 'hologram'. This notion of imaging using interferometric principles laid the foundations for a subsequent revolution in holography, using a variety of wave phenomena including electromagnetic radiation, acoustic waves, and others. Although Gabor's original interpretation of holography was largely restricted to a single-wavelength, it has since been extended to accommodate multiple wavelengths, and in the process ushered a revolution in high-accuracy optical metrology \cite{GaborScience, Yamagiwa:18, KimCombs, Ye:04, JinCombs}.

\setlength{\parindent}{2ex}

The second innovation in Gabor's pioneering work was an \textit{Analysis}/\textit{Synthesis} paradigm that combined wavefront acquisition (\textit{Analysis}) at a smaller wavelength with wavefront correction/reconstruction (\textit{Synthesis}) at a larger wavelength. Gabor utilized this idea to correct for uncompensated spherical aberration in his electron wavelength holograms, using optical lenses designed for visible wavelengths~\cite{GaborAnaSynt}. His approach to optical aberration correction has since been replaced by digital wavefront correction, but the  notion  of  holographic \textit{Analysis}-and-\textit{Synthesis} has endured and is used in this paper to provide deeper insights into the fundamental limits of imaging.

\section*{Synthetic Wavelength Holography (SWH)}

The present work builds on Gabor’s holographic principle with the specific goal of imaging under extensive scatter. The connection to Gabor's \textit{Analysis}/\textit{Synthesis} paradigm is detailed below (and illustrated in  Fig.~\ref{fig:Fig_AH}):
\begin{itemize}

\item 

    \textbf{Analysis:} we record optical wavefronts at two closely spaced wavelengths $\lambda_1$ and $\lambda_2$, each of which is susceptible to scattering. The physical process of scattering may be interpreted as an unknown randomized aberration that irreversibly corrupts the phase of the optical fields $E(\lambda_1)$ or $E(\lambda_2)$, destroying
the ability to recover an image of the object.

\item 

    \textbf{Aberration Correction:}  we exploit spectral correlations in the recorded optical fields to computationally assemble a 'Synthetic Wavelength Hologram' (SWH) $E(\Lambda)~=~E(\lambda_1)E^*(\lambda_2)$, whose phase  is virtually impervious to the effects of scattering at the optical wavelengths $\lambda_1,\lambda_2$. It is demonstrated that the SWH encapsulates field information at a 'Synthetic Wavelength' (SWL) 
	$\Lambda = \frac{\lambda_1 \cdot \lambda_2}{|\lambda_1 - \lambda_2|} >> \lambda_{1},\lambda_2$.

\item 
    
    \textbf{Synthesis:} we digitally play back the SWH at the longer SWL $\Lambda$ to uncover object information that cannot be retrieved at the optical wavelength $\lambda_1,\lambda_2$.

\end{itemize}

\noindent
The principal distinction between Gabor's original approach and the one proposed here, lies in the recording of holograms at \textit{multiple wavelengths}, the \textit{computational} compensation of \textit{unknown} aberrations, and the \textit{digital} replay of the recorded hologram.

\section*{Related Work } 

\begin{figure*}[b!]
\includegraphics[width=\linewidth]{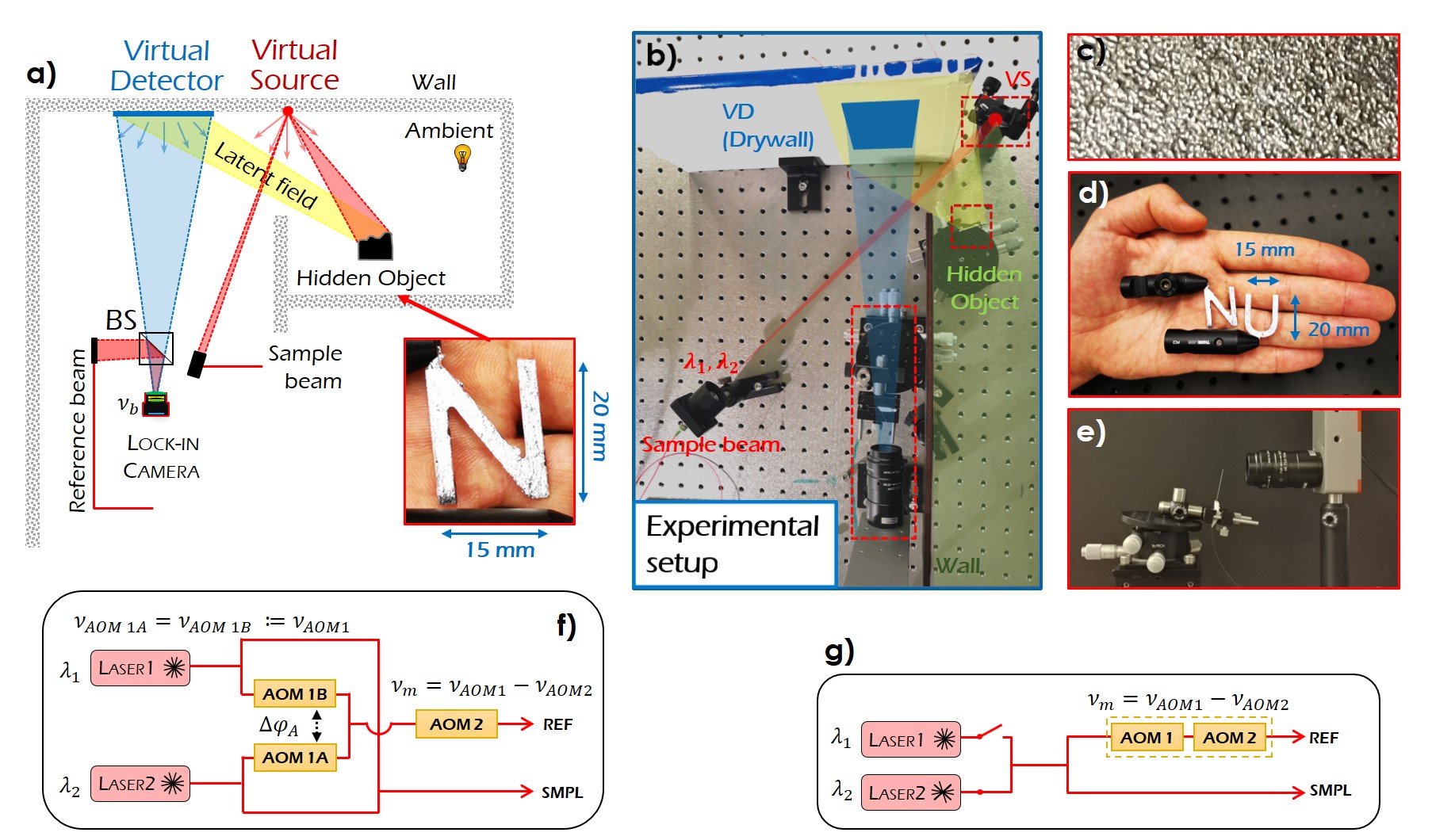}
\caption{Experimental Setup for the `Non-Line-of-Sight' (NLoS) geometry: a) Schematic sketch and image formation: The sample beam illuminates a spot on the wall (the 'Virtual Source' VS), that can be `seen' by the object and the sensor unit. Light is scattered from the VS to the object and from the object surface back to the wall where it hits the `Virtual Detector' (VD). The VD is imaged by the camera, meaning that  the synthetic hologram is captured at the VD surface. b) Picture of the experimental NLoS setup. c) Closeup image of the rough target surface and virtual source (VS) surface: Sandblasted metal coated with silver. d) Image of the used targets: Two characters `N' and `U' with dimensions $\sim 15mm \times 20mm$ (plus black mountings). e) Injection of the reference beam with a `lensed fiber needle' for a minimized light loss.   f) and g)  Interferometer designs used to capture the `Synthetic Wavelegth Hologram' (SWH). Both interferometers introduce a small frequency shift of several $kHz$ between sample  and reference arm, used to demodulate the signal at the SWL. f) Superheterodyne interferometer. g) Dual Wavelength Heterodyne interferometer.}
\label{fig:Fig_B}
\end{figure*}

For numerous tasks in imaging science, information about the object to be imaged is primarily encoded in scattered wavefronts. Classic examples include imaging through densely scattering media such as fog, blood and tissue. A more recent and exciting development is the task of 'Non-Line-of-Sight' (NLoS) imaging - the ability to look around corners using only light that is scattered by a rough wall. The wall serves the dual purpose of indirectly illuminating  obscured objects and intercepting  backscattered light. In the first part of this work, we focus on the NLoS imaging task to motivate and demonstrate the concept of SWH. Later,  we show that our approach can also be applied to image through scattering media. Additionally, we formulate a mathematical framework for analyzing the performance of our method and comparing to other methods for imaging with scattered wavefronts.

While a few passive solutions have been proposed~\cite{saunders19, Lin2020, Bouman, Maeda19, Batarseh18}, the majority of NLoS approaches rely on the availability of an active light source to compensate for significant radiometric losses introduced by scattering from multiple rough surfaces.  Existing work can be broadly categorized into two classes. The first class of techniques (referred to from here on as 'ToF-NLoS') exploits fast (RF-) modulated light sources or short light pulses paired with ultrafast detectors to measure the spatio-temporal impulse response of the obscured scene. Recent publications ~\cite{LiuNat_19, OtooleNat_18, Faccio:19} have  demonstrated NLoS imaging with impressive quality, in some cases providing near real-time reconstructions. The spatial resolution of these methods is presently restricted by a technical limitation: the timing jitter of the source/detector pair. For the commonly used SPAD ('Single Photon Avalanche Diode') detectors, this leads to a maximal spatial resolution in the $cm$ range~\cite{LiuNat_19, OtooleNat_18, Faccio:19}. With a more sophisticated (but very expensive) 'Streak Camera', spatial resolutions under $1cm$ are possible as well~\cite{Velten12}. Since most fast detectors are still limited to single-pixel detection, related approaches rely on raster-scanning, which can cause motion artifacts for moving scenes.
The second class of NLoS techniques exploits spatial/angular correlations in scattered light~\cite{katz14, Aparna_Corr, Balaji:17, PrasannaSPIE}. These techniques recover images of obscured objects at much higher resolution ($\sim 100\mu m$ at $1m$ standoff). However, this  comes at the price of an extremely limited field of view ($<2^\circ$), as determined by the angular decorrelation of scattered light  ('memory effect') \cite{GoodmanSpeck}. 

Interestingly, while a number of high quality NLoS results have been demonstrated in the literature, little attention has been paid to fundamental physical limits of NLoS performance. Current approaches have either demonstrated low resolution imaging over a large field-of-view (FoV), or high resolution over a small FoV. However, the literature remains vague on whether performance can be improved with better hardware, or whether physical limits constrain the maximum FoV and resolution that can be achieved. 
We provide insight into the FOV-resolution tradeoff for imaging with scattered wavefronts, by adapting the Space-Bandwidth Product (SBP) formulation originally developed for conventional imaging. We formulate an upper limit on the SBP and demonstrate experimentally that this limit can be achieved with our SWH method. This is mainly possible because SWH exploits a modality that has rarely been used so far by related approaches: spectral diversity afforded by the use of tunable coherent sources~\cite{ WilloCOSI19, PrasannaSPIE}.

\section*{NLoS Imaging using SWH}

We use the exemplar scene arrangement in Figure~\ref{fig:Fig_B} to elucidate the proposed SWH concept and demonstrate the ability to record holograms of obscured objects. The reflective diffuser designated 'Virtual Source' (VS) in Fig.~\ref{fig:Fig_B}b serves as a proxy for a physical wall that scatters light towards the hidden target. The drywall panel designated 'Virtual Detector' (VD) in Fig.~\ref{fig:Fig_B}b intercepts the light scattered by the target. An imaging optic relays this scattered light onto a focal plane array (FPA) image sensor. The lensed fiber arrangement of Fig.~\ref{fig:Fig_B}e provides the reference beam required for hologram acquisition. The optical field emerging from the VD surface is recorded in a single snapshot by the interferometric setup of Fig.~\ref{fig:Fig_B}f,g (see methods section for details). We interrogate the hidden scene at two closely spaced wavelengths $\lambda_1,\lambda_2$ and record the related fields backscattered from the VD.
Due to scattering at the various surfaces, the recorded holograms  $E(\lambda_1), E(\lambda_2)$ bear no resemblance to a holographic representation of the obscured object (see Fig.~\ref{fig:Fig_AH}). To recover this holographic description, we computationally mix the recorded holograms $E(\lambda_1)E^*(\lambda_2)~=~E(\Lambda)$, as illustrated in Fig.~\ref{fig:Fig_AH}. The resulting synthetic wavelength holograms shown in Figs~\ref{fig:Fig_C}a-e, behave much like a conventional hologram. Consequently, it is possible to digitally replay each hologram for each SWL (see methods section), to reconstitute an image of the hidden object. Results from the process are illustrated in Figs.~\ref{fig:Fig_C}f-j. It is evident that we are able to recover an image of a small character `N' (dimensions $15mm \times 20mm$) despite being obscured from view. 

As it can be seen in Figs.~\ref{fig:Fig_C}f-j, the resolution of the reconstruction improves with decreasing SWL $\Lambda$. This behavior is in complete agreement with results from classical holography. The diffraction limited resolution (minimum resolvable spot radius $\delta x$) of SWH can be quantified using a familiar expression from digital holography:

\begin{figure*}[t]
\centering
\includegraphics[width=0.8\linewidth]{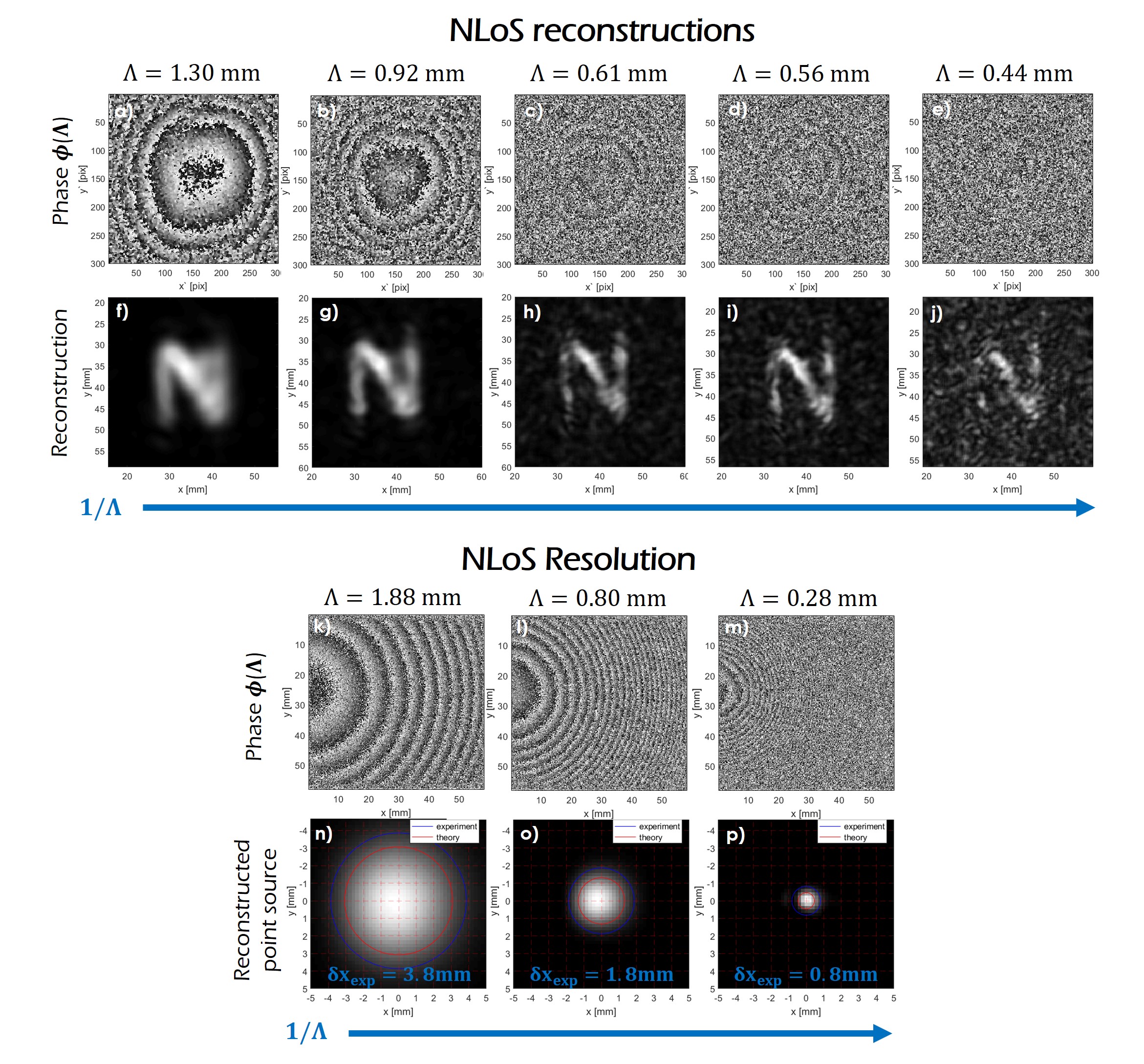}
\caption{Experimental results for NLoS measurements. a)-j) Imaging the character `N' around the corner at five different SWLs. a)-e) Phase maps of  synthetic holograms captured at the VD surface. f)-j) Respective reconstructions. The resolution of the reconstructions increases with decreasing SWL. However, the speckle-artifacts  increase due to the decorrelation of the two optical fields at $\lambda_1$ and $\lambda_2$. k)-p) Reconstruction of a point source around the corner for three different SWLs. k)-m) Phase maps of the synthetic holograms captured at the VD surface. n)-p) Reconstruction of the point source. As in classical optics, the diameter is linearly dependent on the wavelength (in this case the SWL). The experimental value is close to the theoretical expectation. For p), the point source is reconstructed with sub-mm precision. }
\label{fig:Fig_C}
\end{figure*}

\begin{equation}
\delta x \approx \Lambda\frac{z}{D} ~~,
\label{eq:ResLim}	
\end{equation}

\begin{figure*}[b]
\centering
\includegraphics[width=0.6\linewidth]{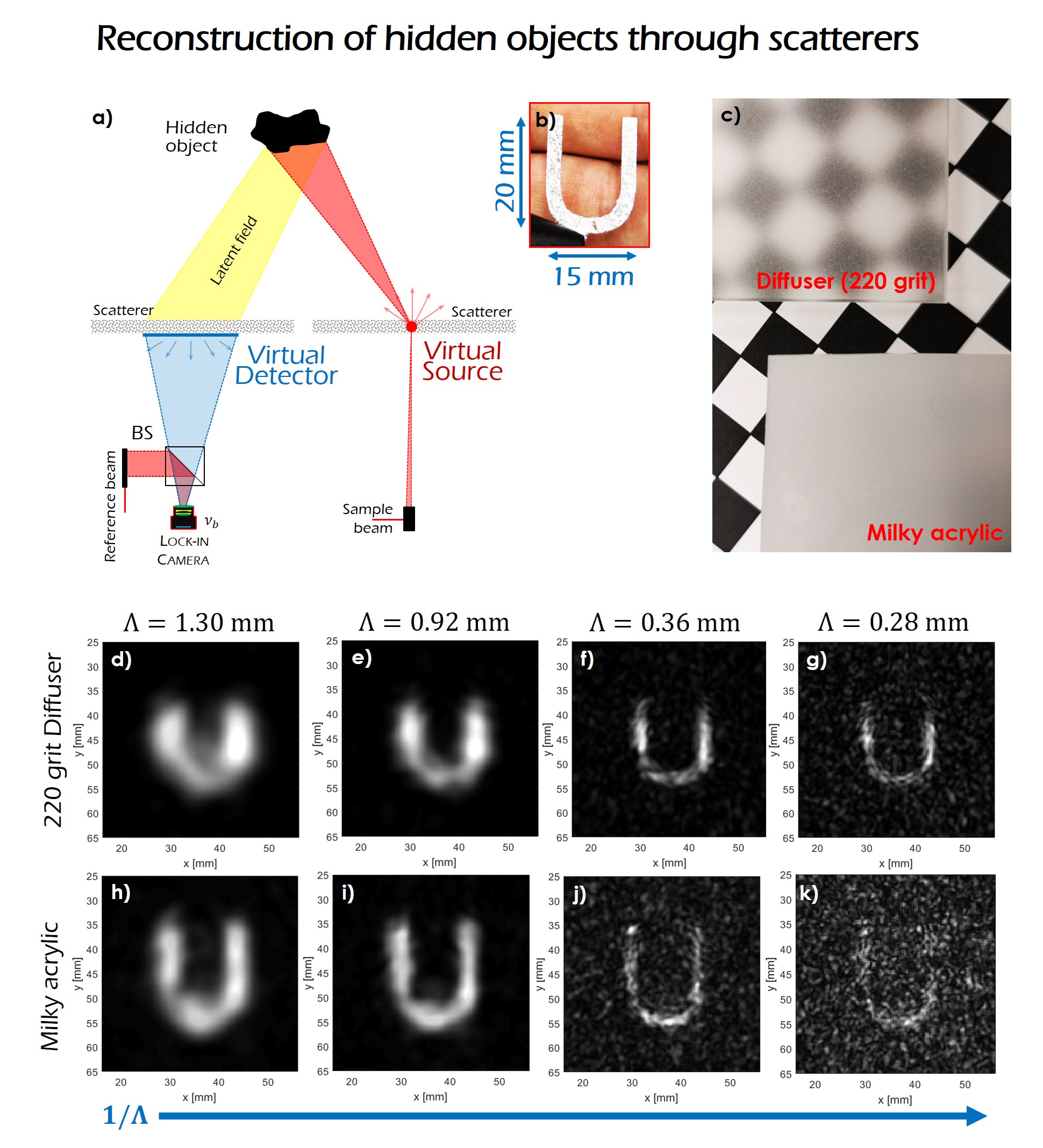}
\caption{Experimental results for measurements through scatterering media. a) Schematic setup. Instead of scattered from a wall, the light is now scattered in transmission. b)~Imaged character `U' with dimensions $\sim 15mm \times 20mm$. c) Scatterers used in the imaging path: A 220 grit ground glass diffuser and a milky plastic acrylic plate of $\sim 4mm$ thickness, both placed $\sim 1cm$ over a checker pattern to demonstrate the decay in visibility. d)-g) Reconstructions of  measurements taken through the ground glass diffuser.  h)-k) Reconstructions of  measurements taken through the milky acrylic plate. The character can be reconstructed with impressive quality. The larger OPD in the acrylic plate leads to a greater decorrelation if the SWL is decreased. }
\label{fig:Fig_D}
\end{figure*}

\noindent
where $D$ is the physical extent of the VD, and  $z$ is the standoff distance between VD and obscured object. Equation~\ref{eq:ResLim} succinctly captures the relationship between the SWL $\Lambda$ and the highest resolution that can be achieved. A smaller SWL is clearly desirable since it leads to higher resolution. We experimentally validate the above claim (and Eq.~\ref{eq:ResLim}) by localizing a point-like source in the hidden volume. An exposed fiber connector positioned $z =95 mm$ behind the VD surface serves as a point-source. Holograms at the VD surface acquired with multiple optical wavelengths are processed to recover a multitude of SWHs, each of which is digitally replayed to recover an image of the point-source. The experimentally observed spot sizes, shown in Figs~\ref{fig:Fig_B}n-p, are consistent with theoretical predictions (red circles, calculated from Eq.~\ref{eq:ResLim} using the  measured VD diameter $D = 58mm$), and increase with increasing SWL. For a SWL of $280 \mu m$, we are able to achieve sub-millimeter resolution around the corner.

\vspace{2mm}

The resolution tests from Figs~\ref{fig:Fig_B}n-p seem to indicate that diffraction limited resolution can be increased indefinitely by decreasing the synthetic wavelength. However, the results in Figs~\ref{fig:Fig_B}f-j demonstrate that for decreasing values of $\Lambda$, the reconstructed image is corrupted with speckle-like artifacts. This suggests a limit to improving the resolving power of SWH. For the moment, we relate this fact as an empirical observation: the SWH is riddled with artifacts when the speckle patterns in the holograms recorded at $\lambda_1$ and $\lambda_2$ are decorrelated. The physical origins of this decorrelation can be traced back to the number of scattering events and the severity of scattering at the VS and VD surfaces (see Supplementary material). 

Closer inspection of the results shown in Figs~\ref{fig:Fig_C}i and~\ref{fig:Fig_C}p reveals a discrepancy between the smallest achievable SWL for the NLoS imaging experiments with the extended object 'N' and the NLoS point-source localization experiment. This discrepancy stems from the fact that light experiences two additional scattering events in the experiments with the character 'N'.  In a later section we provide a mathematical description of this phenomena, one that yields to an insightful and intuitive perspective on the fundamental limits of imaging with scattered wavefronts.

\section*{SWH in Transmissive Scattering Regimes}

The notion of exploiting spectral correlations in scattered light for the purposes of imaging, is by no means restricted to the NLoS problem. To highlight the versatility of the SWH approach, we recover holograms of objects hidden behind a scattering medium, as illustrated in the schematic of Fig.~\ref{fig:Fig_D}a. In a first set of measurements, we image the small character `U' (dimensions $15mm \times 20mm$) through a 220 grit diffuser (Fig.~\ref{fig:Fig_D}c top). The holographic reconstructions of the  character `U' are displayed in Fig.~\ref{fig:Fig_D}d-g. As the SWL approaches  $ 250\mu m$, we begin to notice speckle-like artifacts in the reconstructed image. This leads us to conclude that the separation in optical wavelengths has increased to the point that the captured holograms are no longer correlated for this specific scene.

In a second set of experiments, we swap the diffuser in the imaging path with a $4mm$ thick milky acrylic plate. The consequence of volumetric scatter is made apparent in Figure~\ref{fig:Fig_D}c by comparing the degraded visibility of a checkerboard that is viewed through the acrylic plate and the  220 grit ground glass diffuser. Despite pronounced  scattering in the acrylic plate, we are able to reconstruct the character `U' for SWLs exceeding $360\mu m$, as shown in Figs.~\ref{fig:Fig_D}h-k. This suggests the ability to recover image information at visibility levels far below the perceptual threshold. 
However, a comparison of the reconstructions for the acrylic plate and the diffuser reveals only a marginal change in the smallest achievable SWL. We show that this remarkable observation may be traced to the fact that visibility of ballistic light paths decays exponentially with the propagation distance through a scattering volume (in accordance with Beer’s law~\cite{Beer}), whereas SWH resolution expressed by Eq.~\ref{eq:ResLim} is linearly related to the choice of $\Lambda$.
\begin{figure*}[t]
\centering
\includegraphics[width=\linewidth]{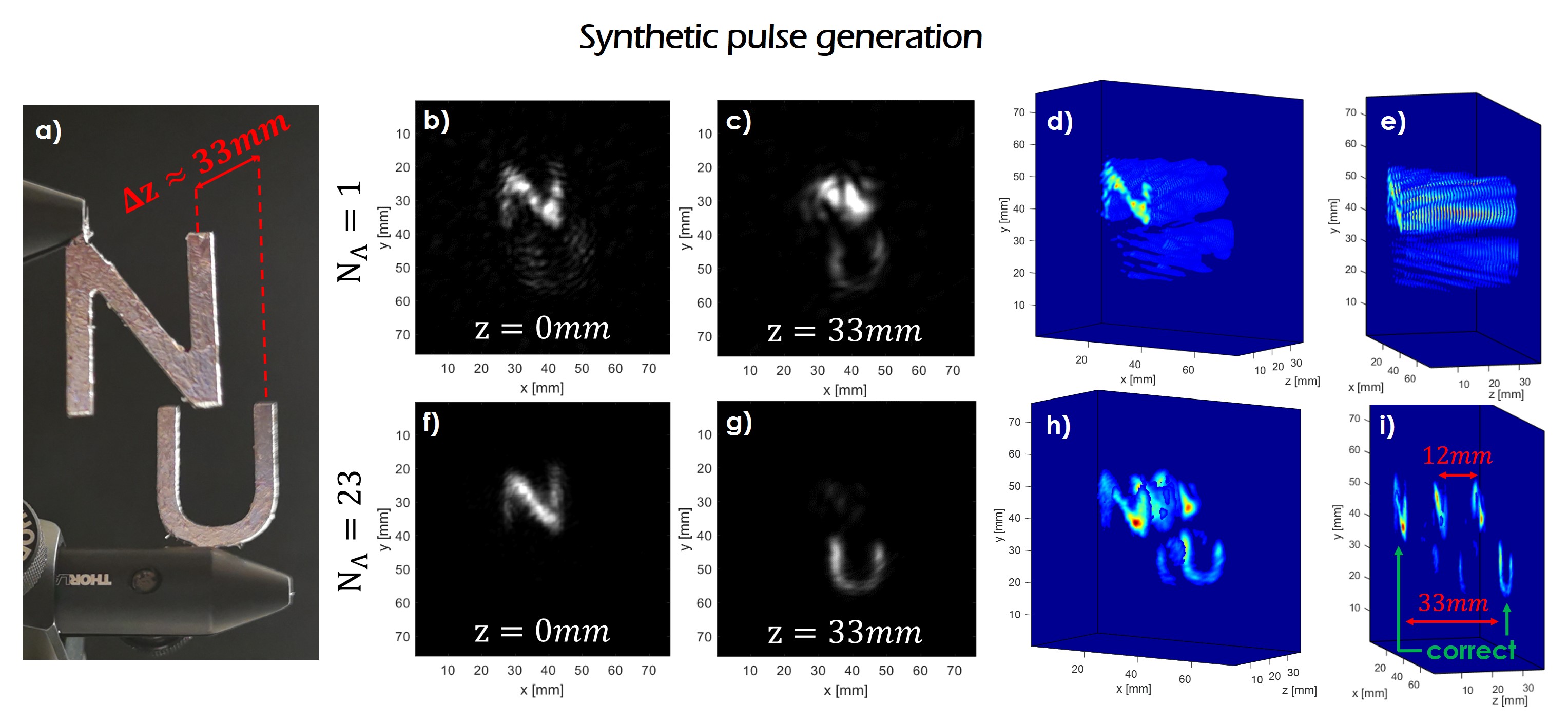}
\caption{Depth separation of two hidden objects by creating a 'synthetic pulse train'. a) Target, consisting of two characters with a longitudinal separation of $33mm$. b)-e) Reconstruction of the characters, using only $N_{\Lambda} = 1$ SWL ($\Lambda=0.8mm$). Due to the properties of holographic backpropagation, a separation of the characters in depth is not possible. f)-i) Reconstruction, calculated from coherent superposition of the backpropagated fields at $N_{\Lambda} = 1$ SWLs. Letters are separable. The pulse distance of the synthesized pulse train can be seen in (h) and (i). }
\label{fig:Fig_E}
\end{figure*} 

In Fig.~\ref{fig:Fig_E}, we demonstrate that the principles underlying the proposed SWH concept are by no means restricted to the use of two wavelengths. The spectral diversity afforded using multiple illumination wavelengths is expected to yield an improvement in the longitudinal resolution, in much the same manner as Optical Coherence Tomography (OCT) \cite{Wojtkowski:10, Andretzky99, FercherOCT, Huang1178} and White-Light Interferometry (WLI) \cite{Leith:80, Dresel:92}. However, unlike OCT and WLI, we neither need to match the pathlengths nor the power in the two arms of our inteferometeric imager (our approach  most closely resembles work by Erons et al. \cite{Arons:93} in Fourier Synthesis Holography). To demonstrate the improvement in longitudinal resolution afforded by the use of multiple SWLs, we computationally section a multi-planar scene consisting of two characters `N' and `U' (introduced in previous experiments) that are offset in depth by $\Delta z\approx33mm$. Using a single SWL of $\Lambda=800\mu m$ it is possible to separate the characters laterally, but with limited longitudinal resolution, as shown in Figs~\ref{fig:Fig_E}b-e. The improved longitudinal resolution is achieved by coherently combining the SWHs recorded at 23 SWLs. The process mimics scene interrogation by a periodic pulse train, and the  replicas observed in the reconstructions of Fig.~\ref{fig:Fig_E}h and~\ref{fig:Fig_E}i are consistent with the periodicity of the computationally engineered pulse train (smallest used frequency offset of $25 GHz$ relates to $12mm$).
An unambiguous measurement range in excess of $33mm$ requires a frequency increment of $\sim 1 GHz$, which has been experimentally verified with our laser system as well. It is anticipated that locking the tunable laser source to a frequency ruler such as a frequency comb will provide improved longitudinal resolution due to the precise phase relationship between the individual comb teeth \cite{Ye:04, KimCombs, JinCombs}. 

The experiments in SWH described thus far have restricted attention to recovering objects obscured by scattering media. However, the principle underlying SWH, namely spectral correlations in scattered light, is rather general and has broader appeal. As an example, we demonstrate the ability to recover residual phase variations in the wavefronts emerging from a volumetric scattering sample. Details of the experimental apparatus are available in Kadobianskyi et.al.~\cite{kadobianskyi18}. The authors of~\cite{kadobianskyi18} recorded speckle fields emerging from $360 \mu m$ and $720 \mu m$ thick scattering samples with a scattering mean free path of $90 \mu m$. In each case, the sample is interrogated by a quasi-monochromatic collimated beam at 801 equally spaced wavelength steps spanning the range $690 nm$ to $940 nm$. By computationally mixing speckle holograms recorded at adjacent wavelengths, we are able to identify a hologram at the SWL of $2.1mm$. Results from the experiment are tabulated in Section 4 of the supplementary material. The phase of the SWH exhibits a distinct spatial structure that is consistent with the observation of interference fringes due to inter-reflections between the laser aperture and a polarized beam splitter in the illumination path; according to the authors of~\cite{kadobianskyi18}.

\section*{SWH and Fundamental Limits of Imaging with Scattered Wavefronts} \label{SBP}

The experimental results presented in the manuscript jointly exploit the expressive power of holography and spectral correlations in scattered light. The resulting SWH approach is a versatile solution to imaging with scattered wavefronts that can accommodate a variety of scales (looking around corners to looking through fog) and scattering mechanisms (multiple surface scatter and volumetric scatter). The approach, however, is not without limitations. In this section, we derive theoretical bounds for SWH that illuminate fundamental limits to NLoS imaging and the broader problem of imaging with scattered wavefronts. Although previous work has alluded to these limits, we are the first to formally describe them in a mathematical framework. Our framework builds upon the  Space-Bandwidth Product (SBP) formulation \cite{Lohmann_89, Lohmann:96, LohmannSanJose} that is frequently used to characterize and bound the performance of a wide variety of imaging modalities \cite{Milojkovic:12, Wagner03, greenbaum2013, WilloDiss_19, Cossairt:11, WilloOE_17}, including holography \cite{Claus_11, Baek:19}.

The SBP reflects a fundamental tradeoff between the FoV~$W$ and the lateral resolution ${\delta x}^{-1}$ of an imaging modality. It is defined as the dimensionless product $W{\delta x}^{-1}$, representing the number of resolvable spots in the image. For a hologram, the SBP can be additionally described as the product of the physical extent $D$ of the holographic detector and the spatial frequency bandwidth $2\nu_x$, where $\nu_x$ represents the highest resolvable frequency in the hologram. Combining the above definitions yields an unified expression for the SBP of a hologram (defined analogously in the $y$-direction):

\begin{equation}
SBP = \frac{W}{\delta x} = 2 D  \nu_x  ~~
\label{eq:SBPintro}	
\end{equation}

\noindent
The maximal SPB is achieved for the highest spatial frequency in a propagating field which is fundamentally limited by the reciprocal of the wavelength \cite{BornWolf}, so that $max(\nu_x) = \lambda^{-1}$. However, the information embedded within this spatial frequency limit is only preserved when the wavefront aberrations are negligible. For classical imaging systems, Lord Rayleigh \cite{Rayleigh} theorized that the maximum tolerable wavefront error $\Psi_{max}$ cannot exceed one quarter of the optical wavelength.  Rayleigh's view has been repeatedly confirmed by optical designers, and commonly referred to as the 'Rayleigh Quarter Wavelength Rule' (RQWR) \cite{Rayleigh, BornWolf, Mahajan:82}:

\begin{equation}
\Psi_{max} \leq \frac{\lambda}{4}~~
\label{eq:ScattCondLim}	
\end{equation}

\noindent
In the presence of scatter, the maximal wavefront error $\Psi_{max}$ represents the worst case Optical Path Difference (OPD) of the numerous scattered light paths that share a common source location, object location and detector pixel. In view of this definition, it is not surprising that the RQWR is violated by scattering processes at optical wavelengths, such as light bouncing off walls (height fluctuations $\sigma_h\gg\lambda$), and light propagation through scattering media like fog or tissue (thickness $L>$ transport mean free path $\ell^*\gg\lambda$). For surface scattering processes, it can be shown that the spread in path lengths is fundamentally limited by $2\sigma_h$, where $\sigma_h$ represents the RMS surface roughness (see Eq. 40 in supplementary material). For volumetric scattering, the spread in path lengths is given by $2\frac{L^2}{\ell^*}$, where $L$ denotes the thickness of the scattering medium and $\ell^*$ denotes the transport mean free path~\cite{Akkermans, weitzpine} (the factor 2 accommodates round trip propagation through the scattering medium).

Our experiments corroborate the claim  that phase information at scales comparable to the SWL is preserved provided that  $\Lambda$ fulfills the RQWR requirement of Eq.~\ref{eq:ScattCondLim}:

\begin{equation}
 \frac{\Lambda}{4} \geq \Psi_{max} \gg \frac{\lambda_{1}}{4},\frac{\lambda_{2}}{4}
\label{eq:ScattCond}	
\end{equation}

\noindent
This means that the synthetic wave, although a computational construct, has distinct characteristics that it shares with a physical wave at the respective wavelength $\Lambda$. 
We validate this observation by drawing attention to the experimental results shown in Fig.~\ref{fig:Fig_D}d-g: Given the (known) surface roughness of the 220 grit diffuser and the geometry of our setup, we estimate the maximal wavefront abberration for this experiment to $\Psi_{max}\approx 65 \mu m$. Speckle-like artifacts start to arise as the SWL $\Lambda$ approaches $4\Psi_{max}$, which is in complete agreement with the RQWR of Eq.~\ref{eq:ScattCond}. The simplicity of the RQWR outlined in Eq.~\ref{eq:ScattCond} is remarkable given the mathematical complexity of analyzing spectral correlations in light scattered by a disordered medium. The existence of such correlations is well documented \cite{Fercher:85, Vry:86, Dandliker:88, GoodmanSpeck, Ruffing:85, Belmonte:10, Shapiro, Genack90, Boer, Akkermans, Mikhailovskaya, Thompson:97, Albada} from a theoretical standpoint, albeit in the ensemble sense.  Experiments demonstrating spectral correlation for a single realization of disorder are available in \cite{vanBeijnum:11, Andreoli15, kadobianskyi18}.
The supplementary material puts forth mathematical arguments supporting the existence of RQWR (Eq.~\ref{eq:ScattCond}) for a single realization of a surface scattering process (see Section 1.6). 
The derivation assumes that \textit{the change in optical path length induced by a small change in the optical frequency is  small }for ray paths that share a common source location, object location and detector pixel. The argument may be extended to accommodate volumetric disorder by adopting a diffusive approach to light propagation~\cite{Thompson:97}.

The relevance of the RQWR (Eq.~\ref{eq:ScattCond}) to imaging in the presence of scatter, emerges in its ability to define the smallest physical and synthetic wavelength that is unaffected by scattering. As stated previously, the synthetic wavelength hologram exhibits speckle artifacts when the RQWR   of Eq.~\ref{eq:ScattCond} is violated. Consequently, we can  relate the SBP of the SWH to its maximum spatial frequency:

\begin{equation}
\nu_x = \frac{1}{\Lambda} \leq \frac{1}{4 ~\Psi_{max}}
\label{eq:FreqLim}	
\end{equation}

\noindent
Incorporating Eq.~\ref{eq:FreqLim} into the definition of the SBP in Eq.~\ref{eq:SBPintro} yields an upper bound on the the maximum SBP that can be achieved:
\begin{equation}
\boxed{SBP ~= ~\frac{W}{\delta x} ~\leq~ \frac{D}{2~\Psi_{max}}}~~.
\label{eq:SBP-lim}	
\end{equation}

\noindent
This bound is well known for physical waves. Here,  we observe that it applies equally well to computationally assembled waves such as in SWH, and Phasor Fields~\cite{LiuNat_19, Reza2:19}. Eq.~\ref{eq:SBP-lim} represents an uncertainty relation that is intrinsic to imaging with scattered wavefronts. It captures the tradeoff between the achievable FoV $W$ and lateral resolution ${\delta x}^{-1}$. We postulate that, although this limit was derived for imaging using SWH, it also \textit{represents a fundamental limit on the maximum SBP that can be attained by any scheme for imaging with scattered wavefronts that obeys the laws of linear optics.} Methods operating close to this limit are as good as physics allows and cannot be improved by the use of better hardware. We clarify this claim by noting that current ToF-NLoS approaches are also subject to the SBP limit of Eq.~\ref{eq:SBP-lim}. The limit cannot be overcome by using shorter pulses (femtosecond) and faster detectors (streak cameras). The optical frequencies that make up the ultrashort pulses will dephase/decorrelate causing dispersion of the pulse following a scattering event. This decorrelation will limit the performance of ToF-NLoS approaches in much the same manner as observed in SWH.

\section*{Discussion and Conclusion} 

The Principle of 'Synthetic Wavelength Holography' introduced in this manuscript  is inspired by Gabor’s original principle for wavefront-based \textit{Analysis}, \textit{Synthesis}, and \textit{Correction}. We studied fundamental limits in imaging performance through densely scattering media, and provided experimental demonstration of SWH reconstructions. We used tunable lasers to demonstrate that our method is able to reach the physical limit of imaging performance for a broad range of scattering conditions, by tuning the SWL to the smallest possible value that does not violate the RQWR. While the experiments in this paper were carried out with baseband frequencies in the optical domain ($100s$ of THz), lock-in detection of our synthetic wavefront is performed at an RF modulation frequency (a few kHz, see methods section). This enables full-field SWH detection without the need for raster-scanning, using state-of-the-art focal plane array cameras. The benefit of this approach has not been discussed in detail, due to our focus on SBP limits. If, however, it is desired to optimize the ’Space-Time-Bandwidth-Product’ (STBP), or the Channel Capacity \cite{Wagner03, WilloDiss_19}, then fast full-field acquisition of our SWH implementation is of high value.

SWH has a broad range of applications including imaging through scattering and turbid media, imaging through obscurants such as fog and smoke, and NLoS imaging.  However, the scale of wavefront error can vary substantially depending on the imaging task. For instance, the typical wavefront error $\Psi_{max}$ in surface scattering processes in NLoS imaging \textit{is below 1 millimeter}, whereas it can be \textit{several centimeters} for imaging through tissue, and \textit{many meters}, for imaging through fog (depending on the transport mean free path $\ell^*$).

The experimental results in Fig.~\ref{fig:Fig_D} clearly demonstrate that our method is able to image through transmissive,  scattering media, even when the visibility at the baseband frequency is extremely poor. 
However, the approach in its present form is best suited for imaging through thin scattering media ($L>\ell^*$), such as the acrylic plate.  For thick media ($L\gg\ell^*$), the increased spread in path lengths (i.e. spread in travel times) severely limits the achievable SWL~\cite{webster03}. The problem may be mitigated by restricting attention to scattered light paths with a prescribed time of travel. 
A specific embodiment of LiDAR that exploits frequency diversity within the detector integration time (FMCW LiDAR~\cite{zhengBook}), is perfectly suited for the task at hand. By combining the time-gating ability of FMCW LiDAR with SWL principle, it may be possible to see through densely scattering media, using a smaller SWL than is otherwise possible. The notion is expected to have important
implications for imaging through participating media such
as fog, clouds, and rain, a problem of particular importance
to Naval surveillance applications, geospatial imaging,
climatology research, tissue imaging or imaging deeper into the brain.

The SWH principle described in this paper digressed slightly from Gabor’s original principle because we focused on the problem of correcting unknown or random wavefront aberrations caused by the scattering of light. However, we also envision a scenario where SWH could be used to compensate for aberrations at the SWL, in a manner that is analogous to the use of adaptive optics in astronomical telescopes. In this scenario, wavefront distortions relative to the SWL are measured using a separate wavefront sensing device observing a guide star (or some other known reference), then the aberrations present in a captured SWH image are corrected in post-processing. This would relax the  Rayleigh Quarter Wave constraint for the SWL expressed in Eq.~\ref{eq:ScattCond}, provided that the wavefront aberration can be measured to within this tolerance.

Gabor’s initial demonstration of optical holography served as a launchpad for subsequent demonstrations of holography using other wave phenomena. We envision our initial demonstrations of optical SWH as a first step in demonstrating a more general solution to the problem of aberration corrected imaging using wavefronts of any physical nature. In particular, our method provides the greatest benefit when signal contrast at baseband frequencies is essential, yet the visibility of this contrast is effectively eliminated  by scattering in a disordered medium. While we have demonstrated SWH with optical baseband frequencies in this paper, we envision that the same principle may also be applied using wavefront sensing of entirely different phenomena. For instance, we envision the possibility of applying the SWH principle to the problem of ultrasound imaging of biological features embedded within deep layers of tissue or coherent X-ray diffraction imaging of specimens embedded in thick, inhomogeneous samples.  We also imagine that the same method could be used to exploit radio antennae arrays (e.g., the VLA) for space-based astronomical imaging at micro and radio frequencies through dense atmosphere, and possibly below the surface of a planet for remote geological exploration.

\newpage

\section*{Methods}

\subsection*{Aberration correction by formation of a SWH}

The aberration correction step adopted in SWH draws inspiration from multi-wavelength interferometry on rough surfaces \cite{Fercher:85, Vry:86, Dandliker:88}. The process illustrated in the right half of  Fig.~\ref{fig:Fig_AH} (purple box) involves recording speckle fields $E(\lambda_1),E(\lambda_2)$ at two closely spaced illumination wavelengths. Due to the stochastic nature of light scattering, the phase $\phi(\lambda_1),\phi(\lambda_2)$ of each field separately is completely randomized and bears no resemblance to the macroscopic structure of the  object. If however, the illumination beams at the two wavelengths originate from the same source position (such as from a single fiber) and the  inhomogeneities in the scattering medium are quasi-static, then the fields incident on the detector are highly correlated. This is because the light at the two wavelengths \textit{traverses nearly identical ray paths and experiences nearly identical path length fluctuations}. This assumption and observation forms the basis of our computational approach to accommodating scatter where we correlate the complex-valued fields to recover the SWH $E(\Lambda) = E(\lambda_1) E^*(\lambda_2)$, with $\Lambda =\frac{\lambda_1\lambda_2}{|\lambda_1-\lambda_2|}$. It can be shown (see supplementary material) that the residual phase fluctuations in the SWH, given by $\phi(\Lambda)= \phi(\lambda_1) - \phi(\lambda_2)$, preserves phase variations at scales equal or larger than the SWL $\Lambda$, and is robust to speckle artifacts. However, the magnitude of the SWH, given by $|E(\Lambda)|=|E(\lambda_1)|\cdot |E(\lambda_2)|$, still exhibits speckle artifacts (see Fig.~\ref{fig:Fig_AH}).

\subsection*{Interferometer design and lock-in detection of the SWH}

The discussion on SBP limits in previous sections has implicitly assumed the availability of idealized sources and detectors. In practice, poor signal-to-background or signal-to-noise ratios, or both, can limit our ability to achieve the theoretical SBP. Interferometric approaches exploiting \textit{frequency heterodyning} have particularly advantageous properties with respect to this problem. The principal benefit of adopting these approaches to record holograms is the ability to exploit the heterodyne gain~\cite{Liu:16} afforded by the use of a strong reference beam, whose baseband optical frequency is slightly detuned from the frequency of light in the object arm. The difference in frequency $\nu_m$ is chosen in the RF frequency range ($3 kHz$ for our experiments) and realized by using a cascade of acousto-optic or electro-optic modulators (AOM or EOM). Figs.~\ref{fig:Fig_B}(f,g) depict the two interferometer designs that we use to acquire the holograms at the two optical wavelengths. Each design is an adaptation of a Michelson Interferometer, and incorporates a small difference $\nu_m$ in the baseband frequency of light in the two arms of the interferometer. It is emphasized that the RF modulation frequency $\nu_m$ is \textit{fully decoupled} from the choice of SWL (and therefore from the resolution of our method!), and can be chosen independent of the SWL.

A Lock-In Focal Plane Array (LI-FPA)~\cite{Heliotis} capable of synchronously demodulating the received irradiance at each detector pixel, is operated to detect the RF frequency $\nu_m$. The process directly yields the interferogram at the SWL  $\Lambda$. The method avoids the need for time consuming raster scanning as necessary in ToF-NLoS, and phase-shifting in holographic-NLoS. It also vastly improves the Signal-to-Background ratio of our measurements by suppressing the unmodulated ambient illumination. The Heliotis C3 LI-FPA~\cite{Heliotis} used in our experiments yields a $300\times300~pix$ image per measurement. The exposure time of each measurement is $t_{exp} = 23 ms$ corresponding to 70 cycles of the RF frequency $\nu_m=3 kHz$.
Two independently tunable narrow linewidth CW lasers (Toptica DFB pro $855nm$) are used to illuminate and interrogate the scene. The center wavelength of each laser is $855 nm$, and the maximum tuning range is $\sim2.5 nm$. This allows us to achieve SWLs $\Lambda > 300 \mu m $, corresponding to a beat frequencies $< 1 THz$. \\

The holograms in our proof-of-principle experiments were recorded using two specific heterodyne interferometer architectures: a Dual-Wavelength Heterodyne Interferometer (Fig.~\ref{fig:Fig_B} g), and a Superheterodyne Interferometer (Fig.~\ref{fig:Fig_B} f). The Dual-Wavelength Heterodyne Interferometer is preferred when light loss in the interferometer should be minimized, which is important for many NLoS applications. Light from the two lasers operating at $\lambda_1,\lambda_2$ are coupled together, before being split into the reference and sample arm. The reference arm is additionally modulated by $\nu_m = 3 kHz$, using a cascade of two fiber AOM's. During acquisition, each laser is shuttered independently and the lock-in camera records the holograms by the two wavelengths, in a time-sequential manner. The LI-FPA provides two images: In-Phase (I) and Quadrature (Q), each of which represents the real and imaginary parts of the speckle fields incident on the image sensor. The expression for the I- and Q-images recorded by the LI-FPA for the wavelength $\lambda_n$ is:

\begin{equation}
\begin{split}
I_I(\lambda_n) =& A_n  \cos(\phi(\lambda_n)) \\
I_Q(\lambda_n) =& A_n \sin(\phi(\lambda_n))~,
\end{split}
\label{eq:Het_IQ}	
\end{equation}

\noindent
where $A_n$ is the amplitude at $\lambda_n$ and $\phi(\lambda_n)$ is the difference in the phase of light in the object and reference arms. Please note that Eq.~\ref{eq:Het_IQ} omits any reference to spatial locations, in the interest of clarity.

Subsequently, the SWH $E(\Lambda)$ is assembled as follows:

\begin{equation}
\begin{split}
E(\Lambda) =& [I_I(\lambda_1) + i I_Q(\lambda_1)] \cdot [I_I(\lambda_2) + i \cdot I_Q(\lambda_2)]^{*} \\
=& A_1 A_2  \exp(i \underbrace{(\phi(\lambda_1)-\phi(\lambda_2))}_{\varphi(\Lambda)})
\end{split}
\end{equation}

\noindent
An attractive feature of the time-sequential approach to hologram acquisition described above is that it does not require the use of two tunable lasers. Identical results can be achieved with one laser that is tuned between the two measurements. Possible extensions include: one tunable and one fixed wavelength laser, and one fixed wavelength laser that is split in two arms, one of which includes an additional modulator.

Unfortunately, the simplicity of the time-sequential approach comes at the expense of increased sensitivity to object motion between measurements, and time-varying fluctuations in the environmental conditions. Increased robustness to these fluctuations is afforded by the Superheterodyne Interferometer design, wherein light from both lasers is used to simultaneously illuminate the target and scene. A possible realization is shown in Fig.~\ref{fig:Fig_B} f: each laser beam is split into two arms, each of which is independently modulated with an AOM. The RF drive frequencies for AOMs 1A and 1B are identically set to $\nu_{AOM1}$, but include a phase offset $\Delta\varphi_{AOM}$ that is user controlled. Light from the two AOMs is combined and modulated with a third AOM (frequency $\nu_{AOM2}$), which produces the desired modulation frequency $\nu_m = \nu_{AOM1} - \nu_{AOM2} = 3kHz$. The expression for the I- and Q-images (In-Phase and Quadrature) recorded by the LI-FPA are:

\begin{equation}
\begin{split}
    I_I(\lambda_1, \lambda_2) =& A_1 \cos(\phi(\lambda_1) + \Delta \varphi_{AOM}) + A_2  \cos(\phi(\lambda_2)) \\
    I_Q(\lambda_1, \lambda_2) =& A_1 \sin(\phi(\lambda_1) + \Delta \varphi_{AOM}) + A_2  \sin(\phi(\lambda_2))
\end{split}
\end{equation}

\noindent
The SWH $E(\Lambda)$ is assembled by calculating:

\begin{equation}
\begin{split}
I_I^2 + I_Q^2 \\
=& A_1^2 + A_2^2 + A_1 A_2  \cos(\underbrace{\varphi(\lambda_1)-\varphi(\lambda_2)}_{\varphi(\Lambda)}) + \Delta \varphi_{AOM})
\end{split}
\end{equation}

\noindent
The synthetic phase map is recovered from the interferograms recorded with three or more phase shifts $\Delta \varphi_{AOM}$ introduced between measurements. It should be emphasized that the use of two tunable lasers is also not a pre-requisite for the approach. Identical results can be achieved with one fixed and one tuned laser, or similar combinations discussed above.
The principal benefit of the Superheterodyne approach is the robustness to environmental fluctuations and object motion. However, it requires an additional AOM and fiber splitters that significantly reduce the available output power compared to the Dual Wavelength Heterodyne Interferometer discussed previously. The loss of power presents light throughput challenges for NLoS experiments that are intrinsically light starved. 

In practice, there exists a trade-off between light throughput and robustness to environmental fluctuations, which depends on a multiple factors including stand-off distance, reflectivity of the involved surfaces, and laser power.

\subsection*{Reference beam injection with reduced radiometric losses}

The reference beam required for interferometric sensing of the speckle fields at the optical wavelengths is directed towards the Lock-In FPA, as shown in Fig.~\ref{fig:Fig_B} a. In one possible embodiment, a \textit{lensed fiber needle} (WT\&T Inc.) positioned in the front focal plane of the imaging optic (see Fig.~\ref{fig:Fig_B} e) produces a near planar reference beam on the FPA. The use of a lensed fiber provides two distinct advantages over a beam-splitter: (1) the imaging optic can be directly threaded to the camera (eliminates the need for inserting beam splitter between optic and sensor) and easily swapped during operation, and (2) improved light throughput (see Tab.~\ref{tab:LLNeedle}).

\begin{table}[h]
	\centering
		\begin{tabular}{ c c c }
\textbf{Light Loss in: } & Reference Beam & Sample Beam \\ 
 Lensed Fiber Needle & $\sim 30\%$ & $\sim 0\%$ \\  
 50/50 Beam Splitter & $\sim 50\%$ & $\sim 50\%$   
		\end{tabular}
	\caption{Light loss at combination of reference and sample arm: Lensed fiber needle vs. conventional 50/50 beam splitter}
	\label{tab:LLNeedle}
\end{table}

\subsection*{Experimental setup  and image formation in NLoS application}

The experimental apparatus of Fig.~\ref{fig:Fig_B} is used to demonstrate the ability of SWH to discern objects obscured from view, in this case a cutout of the character `N' with dimensions $\sim 20mm \times 15mm$. The size of the object was deliberately chosen to be smaller than the typical size of a resolution cell ($\sim 2 cm$) in competing wide-field ToF-NLoS approaches. The disadvantage when using a small object is that it emits less light than the background. The problem is additionally compounded by the limited laser power in the object arm (about $30mW$). In an effort to bypass these engineering limitations, we glued a thin sheet of silver foil to the sandblasted (280 grit) surface of the object 'N' and repeated the process for the VS surface. An image of the VS surface under ambient light (also representative for the surface of the object 'N') is included in  Fig.~\ref{fig:Fig_B}c. In both cases, we ensured that the fields reflected by these materials are fully developed speckle patterns. The VD wall surface is constructed from a standard dry-wall panel that has been painted white (Beer Eggshell).

Our approach to NLoS imaging relies on the availability of an intermediary scattering surface (such as the wall in Fig.~\ref{fig:Fig_B}c) that serves to indirectly illuminate the obscured target and intercept the light scattered by the target. Accordingly, the intermediary surface may be viewed as a Virtualized Source (VS) of illumination and a Virtualized Detector (VD) for the obscured object.

Laser light from the physical source (at wavelengths $\lambda_1$ and $\lambda_2$) is directed towards the  VS surface using a focusing optic. This light is scattered by the VS surface so as to illuminate the obscured object with a fully developed objective speckle pattern. A fraction of the light incident on the obscured object is redirected towards the VD surface. A second scattering event at the VD surface directs a tiny fraction of the object light towards the collection aperture, and subsequently the LI-FPA. The speckle fields impinging on the LI-FPA are synchronously demodulated to recover the real and imaginary parts of the holograms at the optical wavelengths $\lambda_1$ and $\lambda_2$. Each of these holograms is additionally subject to diffraction due to the finite collection aperture. However, the diffraction effects are observed at optical wavelengths and have little impact on the SWL $\Lambda$. After assembling the SWH, the hidden object can be reconstructed by backpropagating the SWH, using a propagator (Free-Space propagator) at the SWL $\Lambda$.

Figure~\ref{fig:Fig_C} includes the result of processing the NLoS measurements acquired using the experimental setup of Fig.~\ref{fig:Fig_B}. The measurements were captured at different SWLs ranging from $280\mu m$ to $2.6mm$. Figure~\ref{fig:Fig_C} shows five exemplary results for $\Lambda= 1.30mm$, $\Lambda= 920 \mu m$, $\Lambda=610\mu m$, $\Lambda=560 \mu m$ and $\Lambda=440 \mu m$. The phase of the SWH associated with each SWL is shown in Fig.~\ref{fig:Fig_C} a-e. The phasemaps have been low-pass filtered with kernel size $\approx \Lambda$ for better visualization.

As discussed previously, the reconstruction resolution improves with decreasing SWL. However, decreasing the SWL leads to an increased spectral decorrelation of the speckle fields at the two optical wavelengths. The decorrelation manifests as excessive phase fluctuations in the SWH, which in turn produces increased speckle artifacts in the reconstructed images. The problem can be mitigated (to an extent) by exploiting speckle diversity at the VS, specifically by averaging over multiple speckle realizations of the virtualized illumination. In our experiment, we realized the speckle diversity by small movements of the VS position. The image insets in Figure~\ref{fig:Fig_C}f-j represent the result of  incoherent averaging (intensity-averaging) of the backpropagated images, for 5 different VS positions. The improvement in reconstruction quality comes at the expense of increased number of measurements, but not unlike competing ToF-NLoS approaches (e.g. $>20.000$ VS positions are used in~\cite{LiuNat_19}). The distinction is that we need far fewer images.
We conclude our discussion by observing that for static objects, the reconstruction quality may be further improved by increasing the number of VS positions used to realize speckle diversity.

\subsection*{Experimental setup and image formation for imaging through scattering media}

The experimental apparatus of Fig.~\ref{fig:Fig_D}a is used to demonstrate the ability of SWH to image through scattering media. 
In a first experiment, we illuminate and image the character `U' (see Fig.~\ref{fig:Fig_D} b) through an optically rough ground glass diffuser (220 grit). The geometry is unlike other transmission mode experiments  wherein the object is  illuminated directly \cite{Singh_14} or sandwiched between two diffusers. The current choice of geometry is deliberate and designed to mimic the imaging of a  target embedded in a scattering medium. Measurements were acquired for different SWLs ranging from $280\mu m$ to $2.6mm$. Figures~\ref{fig:Fig_D}
d-g show four exemplary reconstructions for $\Lambda= 1.30mm$, $\Lambda= 920 \mu m$, $\Lambda=360\mu m$, and $\Lambda=280 \mu m$. In each instance, we incoherently averaged the reconstruction results for two VS positions. A comparison of the image insets in Figures~\ref{fig:Fig_D} confirms the increased decorrelation for decreasing SWL. As discussed previously, the wavefront error for the diffuser is estimated to be $\Psi \approx 65 \mu m$, and the results for $\Lambda = 280 \mu m$ demonstrate performance close to the physical limit expressed by Eq.~\ref{eq:SBP-lim}.

In a second experiment the ground glass diffuser within the imaging path is swapped with  a milky acrylic plastic plate of $\sim 4mm$ thickness. The acrylic plate exhibits pronounced multiple scattering, representative of imaging through volumetric scatter. Figure~\ref{fig:Fig_D}c compares the visibility of a checkerboard viewed through the 220 grit ground glass diffuser and the acrylic plastic plate. In both cases, the checkerboard is positioned $1cm$ under the scattering plate and viewed under ambient illumination. It is evident from Figure~\ref{fig:Fig_D}c that the visibility of the checkerboard pattern is vastly diminished when viewed through the acrylic plate, whereas the pattern is still visible when viewed through the diffuser.

Figure~\ref{fig:Fig_D}h-k shows reconstruction results for the same character `U' as imaged through the acrylic plate, for the same set of SWLs as the diffuser. In each instance, we incoherently averaged the reconstruction results for two VS positions. The character is reconstructed with high fidelity despite pronounced multiple scattering, suggesting the potential of SWH for imaging through volumetric scatter. A comparison of the image insets in Figures~\ref{fig:Fig_D} confirms the diminished fidelity of imaging through volumetric scattering when compared to surface scatter.\\

\begin{acknowledgements}
\noindent
This work was supported by DARPA through the DARPA REVEAL project (HR0011-16-C-0028), by NSF CAREER (IIS-1453192), and ONR (N00014-15-1-2735).
The authors acknowledge the assistance of Mykola Kadobianskyi for making available the experimental data and supporting MATLAB scripts from~\cite{kadobianskyi18}.
The authors gratefully acknowledge the help of Predrag Milojkovic, Gerd Häusler, Ravi Athale, and Joseph Mait in proofreading the manuscript and providing valuable comments. 

\end{acknowledgements}

\section*{Bibliography}
\bibliography{zHenriquesLab-Mendeley}


\end{document}